# Entropy and weak solutions in the LBGK model

**Zheng Ran and Yupeng Xu**

Shanghai Institute of Applied Mathematics and Mechanics, Shanghai University, Shanghai 200072, China

In this paper, we derive entropy functions whose local equilibria are suitable to recover the Euler-like equations in the framework of the Lattice Boltzmann method. Numerical examples are also given, which are consistent with the above theoretical arguments.

## I. INTRODUCTION

In the recent years, the Lattice Boltzmann Method-the Lattice BGK (LBGK) models have raised a considerable interest for the simulation of complex hydrodynamic phenomena, and have proved successful at simulating nearly incompressible and isothermal fluid flows, for which they are a viable alternative to conventional numerical methods [1, 2, 3, 4]. However, applications to flows with substantial temperature fluctuations have proved problematic. Many schemes are subject to instabilities [5], often attributed to the use of polynomial approximations for the equilibrium distributions. These equilibria may become negative, which violates the assumptions of a discrete H-theorem and may permit instability [6]. This has spurred recent work on constructing non-polynomial equilibria that remain positive, and often arise from extremising some entropy function [7]. To date, there are still several directions where the LBM theory calls for substantial progress and upgrades. One of the most interesting problem is the H theorem in lattice BGK model.

Another interpretation of the lattice Boltzmann approach is as a discretized version of the continuum Boltzmann equation. The governing equations of gas-dynamics are expressions of conservation and the second law of thermodynamics. Conservation requires that three fundamental quantities – mass, momentum, and energy – are neither created nor destroyed but are only redistributed or, excepting mass, converted from one form to another. A companion principle to conservation, known as the second law of thermodynamics, requires that a fourth fundamental quantity called entropy should never decrease. The second law of thermodynamics restricts the redistributions and conversions of conserved quantities otherwise allowed by the conservation laws. As another supplement to conservation, the equations of state, the second law of thermo-dynamics collectively constitute of the Euler equations. An interesting problem we asked: How to realization this counterparts in LBGK model? This is the main topic of this letter. It is believed that this is first study that has laid the theoretical foundation of the LBM for the simulation of flows with shock waves and contact discontinuities.

## II. BGK BOLTZMANN EQUATION AND ITS HYDRODYNAMICS

We begin with the Bhatnagar-Gross-Krook[8] Boltzmann equation, which is a model kinetic equation widely studied

$$\partial_t f + \boldsymbol{\xi} \cdot \nabla f = -\omega(f - g) \tag{1}$$

Where the single particle distribution function $f \equiv f(\mathbf{x}, \boldsymbol{\xi}, t)$ is a time-dependent function of particle coordinate **x** and velocity $\boldsymbol{\xi}$, $\omega$ is the relaxation parameter which characterizes typical collision process, and $g$ is the local Maxwellian equilibrium distribution function defined by

$$g(\rho,\mathbf{u},\theta) = \rho(2\pi\theta)^{-D/2} \exp\left[-(\boldsymbol{\xi}-\mathbf{u})^2/2\theta\right] \quad (2)$$

Where D is the dimension of the velocity space $\boldsymbol{\xi}$, $\rho$, **u** and $\theta = k_B T/m$ are the mass density, macroscopic velocity and normalized temperature per unit mass, respectively. $k_B$, $T$ and $m$ are the Boltzmann constant, temperature and particle mass respectively. The mass density $\rho$, velocity **u** and temperature $\theta$ (or internal energy density) are the hydrodynamic moments of $f$ or $g$ [9]:

$$\rho = \int d\xi\, f = \int d\xi\, g \quad (3)$$

$$\rho\mathbf{u} = \int d\xi\, \xi f = \int d\xi\, \xi g \quad (4)$$

$$\frac{D}{2}\rho\theta = \int d\xi\, \frac{1}{2}(\xi-u)^2 f = \int d\xi\, \frac{1}{2}(\xi-u)^2 g \quad (5)$$

At equilibrium $f = g$, the Boltzmann equation (1.1) becomes

$$D_t g = 0 \quad (6)$$

Where

$$D_t \equiv \partial_t + \boldsymbol{\xi}\cdot\nabla \quad (7)$$

The Euler equation can be easily derived from the above equation by evaluating its hydrodynamics moments

$$\partial_t \rho + \nabla\cdot(\rho\mathbf{u}) = 0 \quad (8)$$

$$\partial_t (\rho\mathbf{u}) + \nabla(\rho\mathbf{u}\mathbf{u} + \rho\theta) = 0 \quad (9)$$

$$\partial_t \left(\frac{D}{2}\rho\theta + \frac{1}{2}\rho u^2\right) + \nabla\cdot\left(\frac{D+2}{2}\rho\theta\mathbf{u} + \frac{1}{2}\rho u^2\mathbf{u}\right) = 0 \quad (10)$$

Where **uu** denotes a second rank tensor $u_\alpha u_\beta$ and $u^2 = \mathbf{u}\cdot\mathbf{u}$.

The momentum equation can also be rewritten as the following

$$\rho\partial_t \mathbf{u} + \rho \mathbf{u} \cdot \nabla \mathbf{u} = -\nabla P \tag{11}$$

where

$$P = \rho\theta \tag{12}$$

is the equation of state for ideal gas.

## III. LATTICE BOLTZMANN EQUATION AND ITS HYDRODYNAMICS

The Lattice Boltzmann equation can be derived by discretizing the continuous BGK equation in both time and phase space. The obtained Lattice Boltzmann equation is

$$f_i(t+1, \mathbf{x}+\mathbf{c}_i) - f_i(t, \mathbf{x}) = -\omega\left[f_i(t, \mathbf{x}) - f_i^{eq}(t, \mathbf{x})\right] \tag{13}$$

$f_i^{eq}$ is the discretized equilibrium function and $\{\mathbf{c}_i\}$ is the discrete velocity set. For the sake of explicitness, we use the nine-bit Lattice Boltzmann equation in Two-dimensional space in the following. the equilibrium distribution function of the nine-bit model is

$$f_i^{eq} = w_i \rho \left[ 1 + \frac{3(\mathbf{c}_i \cdot \mathbf{u})}{c^2} + \frac{9(\mathbf{c}_i \cdot \mathbf{u})^2}{2c^4} - \frac{3u^2}{2c^2} \right] \tag{14}$$

where

$$w_i = \begin{cases} 4/9 & i=0 \\ 1/9 & i=1,2,3,4 \\ 1/36 & i=5,6,7,8 \end{cases} \tag{15}$$

$$c_i = \begin{cases} (0,0) & i=0 \\ (\cos((i-1)\pi/2), \sin((i-1)\pi/2)) & i=1,2,3,4 \\ (\cos((i-5)\pi/2), \sin((i-5)\pi/2)) & i=5,6,7,8 \end{cases} \tag{16}$$

are the associated weight coefficient and discrete velocity set. The hydrodynamics moments of the Lattice Boltzmann equation are given by

$$\rho = \sum_i f_i = \sum_i f_i^{eq} \tag{17}$$

$$\rho \mathbf{u} = \sum_i \mathbf{c}_i f_i = \sum_i \mathbf{c}_i f_i^{eq} \tag{18}$$

$$\frac{D}{2}\rho\theta = \sum_i \frac{1}{2}(\mathbf{c}_i - \mathbf{u})^2 f_i = \sum_i \frac{1}{2}(\mathbf{c}_i - \mathbf{u})^2 f_i^{eq} \tag{19}$$

Through the Chapman-Enskog [10, 11] analysis (APPENDIX),

$$D_t f_i^{(eq)} = -\omega f_i^{(1)} \tag{20}$$

The moments of the zeroth order Eq. (20), in the discrete momentum space, lead to the Euler equations

$$\partial_t \rho + \nabla \cdot (\rho \mathbf{u}) = 0 \tag{21}$$

$$\partial_t (\rho \mathbf{u}) + \nabla (\rho \mathbf{u}\mathbf{u} + \rho\theta) = 0 \tag{22}$$

$$\partial_t \left(\frac{D}{2}\rho\theta + \frac{1}{2}\rho u^2\right) + \nabla \cdot \left(\frac{D+2}{2}\rho\theta\mathbf{u}\right) = 0 \tag{23}$$

Note that Eq. (23) differs from its counterpart, Eq. (10), derived from the continuous BGK Boltzmann equation. (Kun Xu, Lishi Luo, 1999) The energy flux $\frac{1}{2}\rho u^2 \mathbf{u}$ due to the advection of fluid is missing. Thus, shock waves across which the entropy of fluid particles decreases may occur in the numerical results. However, we believe that such shock will not consistent with the real physical solutions.

## IV. THE DERIVATION OF THE ENTROPY EQUATION

The derivation of the entropy equation from the equations of (21)-(23) in the one-dimensional problem as follows. Rewrite equations of (21)-(23) as

$$\partial_t \rho + \partial_x (\rho u) = 0 \tag{24}$$

$$\partial_t (\rho u) + \partial_x (\rho u^2 + P) = 0 \tag{25}$$

$$\partial_t(\rho E)+\partial_x(pu+\rho Eu)=\partial_x\left(\frac{1}{2}\rho u^3\right) \tag{26}$$

where

$$E=e+\frac{1}{2}\rho u^2 \tag{27}$$

Eq. (26)- $E\times$ Eq. (24), we have

$$\rho\partial_t E+\rho u\partial_x E+\partial_x(Pu)=\partial_x\left(\frac{1}{2}\rho u^3\right) \tag{28}$$

Then substituting Eq. (27) into Eq. (28), we have

$$\partial_t e+u\partial_x e=\frac{1}{\rho}\left[\partial_x\left(\frac{1}{2}\rho u^3\right)-\rho\partial_t\left(\frac{1}{2}u^2\right)-\rho u\partial_x\left(\frac{1}{2}u^2\right)-\partial_x(Pu)\right] \tag{29}$$

Based on the definition of entropy

$$dS=dQ/T \tag{30}$$

It is well known that

$$de=dQ+\frac{P}{\rho^2}d\rho \tag{31}$$

Hence

$$TdS=de-\frac{P}{\rho^2}d\rho \tag{32}$$

At this stage, we have

$$\partial_t S+u\partial_x S=\frac{1}{T}\left[(\partial_t e+u\partial_x e)-\frac{P}{\rho^2}(\partial_t\rho+u\partial_x\rho)\right] \tag{33}$$

The substituting Eq. (24) into Eq. (33), we obtain

$$\partial_t S+u\partial_x S=\frac{1}{T}\left(\partial_t e+u\partial_x e+\frac{P}{\rho}\partial_x u\right) \tag{34}$$

At last we could deduce the entropy equation as following

$$\partial_t S + u\partial_x S = \frac{1}{\rho T}\left[\partial_x\left(\frac{1}{2}\rho u^3\right) - \rho\partial_t\left(\frac{1}{2}u^2\right) - \rho u\partial_x\left(\frac{1}{2}u^2\right) - \partial_x(Pu) + P\partial_x u\right]$$

$$= \frac{1}{\rho T}\left[\partial_x\left(\frac{1}{2}\rho u^3\right) - \rho u\partial_t u - \rho u^2\partial_x u - u\partial_x P\right]$$

$$= \frac{1}{\rho T}\left[\partial_x\left(\frac{1}{2}\rho u^3\right) - u(\rho\partial_t u) - \rho u^2\partial_x u - u\partial_x P\right]$$

$$= \frac{1}{\rho T}\left\{\partial_x\left(\frac{1}{2}\rho u^3\right) - u\left[-\partial_x P - \partial_x(\rho u^2) - u\partial_t \rho\right] - \rho u^2\partial_x u - u\partial_x P\right\}$$

$$= \frac{1}{\rho T}\left\{\partial_x\left(\frac{1}{2}\rho u^3\right) - u\left[-\partial_x P - \partial_x(\rho u^2) + u\partial_x(\rho u)\right] - \rho u^2\partial_x u - u\partial_x P\right\}$$

$$= \frac{1}{\rho T}\left\{\partial_x\left(\frac{1}{2}\rho u^3\right) + u\partial_x P + u\partial_x(\rho u^2) - u^2\partial_x(\rho u) - \rho u^2\partial_x u - u\partial_x P\right\}$$

$$= \frac{1}{\rho T}\left\{\partial_x\left(\frac{1}{2}\rho u^3\right) + u\partial_x(\rho u^2) - u^2\partial_x(\rho u) - \rho u^2\partial_x u\right\}$$

$$= \frac{1}{\rho T}\left\{\partial_x\left(\frac{1}{2}\rho u^3\right) + u^2\partial_x(\rho u) + \rho u^2\partial_x u - u^2\partial_x(\rho u) - \rho u^2\partial_x u\right\}$$

$$= \frac{1}{\rho T}\partial_x\left(\frac{1}{2}\rho u^3\right) \tag{35}$$

$$\partial_t S + u\partial_x S = \frac{1}{2\rho T}\left(u^3\partial_x \rho + 3\rho u^2\partial_x u\right) \tag{36}$$

Now we could compute

$$C = u^3\partial_x \rho + 3\rho u^2\partial_x u$$

to demonstrate the entropy preservation condition in the LBGK calculation.

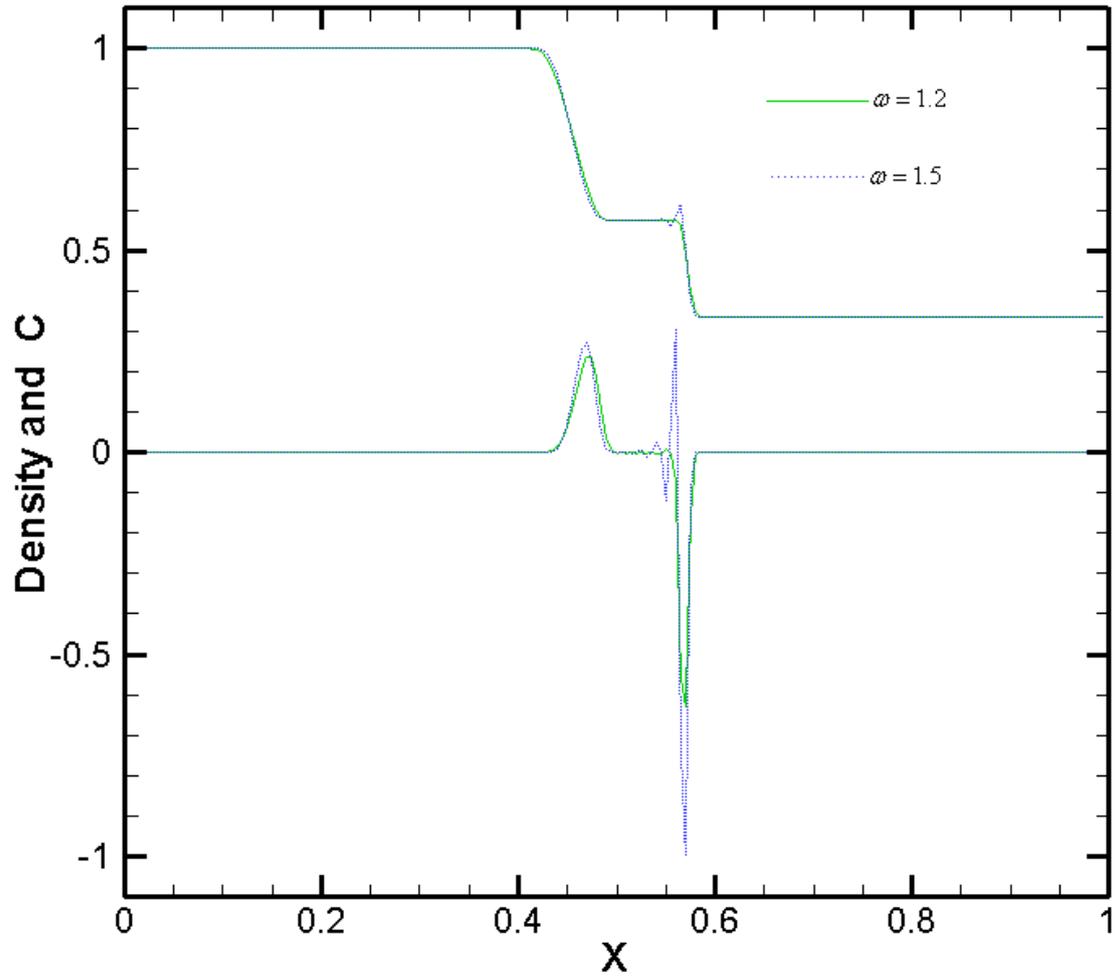

FIG. 1. the density and entropy profiles with different relaxation parameter for D1Q3 model.

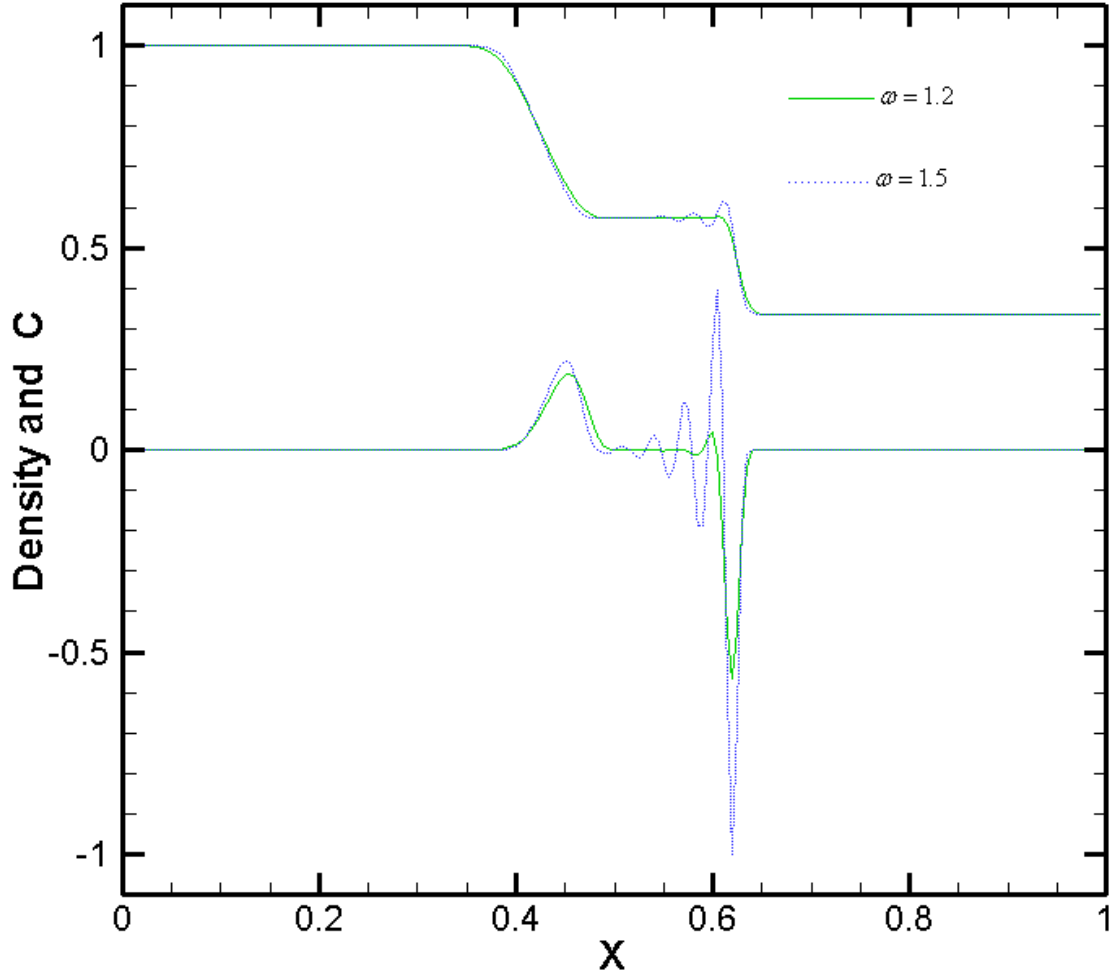

FIG.2. the density and entropy profiles with different relaxation parameter for D1Q5 model.

## V. CONCLUSION

We use the one-dimensional three and five velocity model (D1Q3, D1Q5) to numerically study the entropy in LBGK calculating. In Fig.1 and Fig.2, the entropy profiles are presented at different time for D1Q3 and D1Q5, and different $\omega$ ranging from 1.2 to 1.5. In all cases, we observe a negative entropy range exist near the shock, while oscillatory are captured. We reported in this letter the entropy calculation in LBGK model. Great care has been paid to the links between the oscillation solution and the entropy negative. An explicit relation is obtained and confirmed by the numerical simulation, which must be in accordance with the second law of thermodynamics. The entropy condition for general LBGK method is quite interesting and debatable question.

## Appendix: Chapman-Enskog procedure

(1) Taylor expansion for LBGK equation

The Chapman-Enskog method is the standard procedure used in statistical mechanics to solve an equation like (13) with a perturbation parameter.

$$f_i(t+1, \mathbf{x}+\mathbf{c}_i) - f_i(t, \mathbf{x}) = -\omega \left[ f_i(t, \mathbf{x}) - f_i^{eq}(t, \mathbf{x}) \right]$$

Equation (13) can be Taylor expanded up to second order. A straightforward calculations gives

$$\left( \partial_t + c_{i\alpha} \partial_\alpha \right) f_i + \frac{1}{2} \left( \partial_t + c_{i\alpha} \partial_\alpha \right)^2 f_i = \omega \left( f_i^{(eq)} - f_i \right) \tag{A1}$$

(2) Multiscale expansion

Introduce two time scales $t_1 = \varepsilon t$, $t_2 = \varepsilon^2 t$ and one spatial scale $\mathbf{x}_1 = \varepsilon \mathbf{x}$ and multiscale expanded for time, spatial and $f_i$ [12]

$$\partial_t = \varepsilon \partial_{t_1} + \varepsilon^2 \partial_{t_2}$$

$$\partial_\alpha = \varepsilon \partial_{1\alpha}$$

$$f_i = f_i^{(eq)} + \varepsilon f_i^{(1)} + \varepsilon^2 f_i^{(2)} + \cdots$$

Substituting them into Eq. (A1), we obtain

$$\left( \varepsilon \partial_{t_1} + \varepsilon^2 \partial_{t_2} + \varepsilon c_{i\alpha} \partial_{1\alpha} \right) \left( f_i^{(eq)} + \varepsilon f_i^{(1)} + \varepsilon^2 f_i^{(2)} + \cdots \right)$$

$$+ \frac{1}{2} \left( \varepsilon \partial_{t_1} + \varepsilon^2 \partial_{t_2} + \varepsilon c_{i\alpha} \partial_{1\alpha} \right)^2 \left( f_i^{(eq)} + \varepsilon f_i^{(1)} + \varepsilon^2 f_i^{(2)} + \cdots \right)$$

$$= \omega \left( f_i^{(eq)} - f_i^{(eq)} - \varepsilon f_i^{(1)} - \varepsilon^2 f_i^{(2)} - \cdots \right)$$

$$\varepsilon \left( \partial_{t_1} + c_{i\alpha} \partial_{1\alpha} \right) f_i^{(eq)} + \varepsilon^2 \left[ \partial_{t_2} f_i^{(eq)} + \left( \partial_{t_1} + c_{i\alpha} \partial_{1\alpha} \right) f_i^{(1)} \right]$$

$$+ \varepsilon^2 \frac{1}{2} \left( \partial_{t_1} + c_{i\alpha} \partial_{1\alpha} \right)^2 f_i^{(eq)} + o(\varepsilon^2) = \omega \left( f_i^{(eq)} - f_i^{(eq)} - \varepsilon f_i^{(1)} - \varepsilon^2 f_i^{(2)} - \cdots \right)$$

Comparing the coefficients:

$$\varepsilon^1: \quad \left( \partial_{t_1} + c_{i\alpha} \partial_{1\alpha} \right) f_i^{(eq)} = -\omega f_i^{(1)}$$

$$\varepsilon^2: \quad \partial_{t_2} f_i^{(eq)} + \frac{1}{2} \left( \partial_{t_1} + c_{i\alpha} \partial_{1\alpha} \right)^2 f_i^{(eq)} + \left( \partial_{t_1} + c_{i\alpha} \partial_{1\alpha} \right) f_i^{(1)} = -\omega f_i^{(2)}$$

(3) The derivation of Eq. (23)

For the derivation of Euler equation, only $\varepsilon^1$ scale researched.

$$\left( \partial_{t_1} + c_{i\alpha} \partial_{1\alpha} \right) f_i^{(eq)} = -\omega f_i^{(1)} \tag{A2}$$

Based on Eq. (17), (18), (19)

$$\frac{D}{2} \rho \theta = \sum_i \frac{1}{2} (\mathbf{c}_i - \mathbf{u})^2 f_i = \sum_i \frac{1}{2} (\mathbf{c}_i - \mathbf{u})^2 f_i^{eq}$$

$$\frac{D}{2} \rho \theta = \sum_i \frac{1}{2} c_i^2 f_i - \sum_i u_\alpha c_{i\alpha} f_i + \sum_i \frac{1}{2} u^2 f_i = \sum_i \frac{1}{2} c_i^2 f_i^{eq} - \sum_i u_\alpha c_{i\alpha} f_i^{eq} + \sum_i \frac{1}{2} u^2 f_i^{eq}$$

$$\frac{D}{2} \rho \theta = \sum_i \frac{1}{2} c_i^2 f_i - u_\alpha \sum_i c_{i\alpha} f_i + \frac{1}{2} u^2 \sum_i f_i = \sum_i \frac{1}{2} c_i^2 f_i^{eq} - u_\alpha \sum_i c_{i\alpha} f_i^{eq} + \frac{1}{2} u^2 \sum_i f_i^{eq}$$

$$\frac{D}{2} \rho \theta = \sum_i \frac{1}{2} c_i^2 f_i - u_\alpha \sum_i \mathbf{c}_{i\alpha} f_i + \frac{1}{2} u^2 \sum_i f_i = \sum_i \frac{1}{2} c_i^2 f_i^{eq} - u_\alpha \sum_i c_{i\alpha} f_i^{eq} + \frac{1}{2} u^2 \sum_i f_i^{eq}$$

$$\frac{D}{2} \rho \theta = \sum_i \frac{1}{2} c_i^2 f_i - \rho u^2 + \frac{1}{2} \rho u^2 = \sum_i \frac{1}{2} c_i^2 f_i^{eq} - \rho u^2 + \frac{1}{2} \rho u^2$$

$$\frac{D}{2} \rho \theta = \sum_i \frac{1}{2} c_i^2 f_i - \frac{1}{2} \rho u^2 = \sum_i \frac{1}{2} c_i^2 f_i^{eq} - \frac{1}{2} \rho u^2$$

Then we obtain

$$\frac{D}{2} \rho \theta + \frac{1}{2} \rho u^2 = \sum_i \frac{1}{2} c_i^2 f_i = \sum_i \frac{1}{2} c_i^2 f_i^{eq} \tag{A3}$$

$$0 = \sum_i \frac{1}{2} c_i^2 f_i^{(1)} \tag{A4}$$

Take the D2Q9 model for example, based on Eq. (14)

$$f_i^{eq} = w_i \rho \left[ 1 + \frac{3(\mathbf{c}_i \cdot \mathbf{u})}{c^2} + \frac{9(\mathbf{c}_i \cdot \mathbf{u})^2}{2c^4} - \frac{3u^2}{2c^2} \right]$$

Then the calculation as follow

$$\sum_i c_{i\alpha} c_{i\beta} c_{i\gamma} f_i^{eq}$$

$$= \sum_i c^3 e_{i\alpha} e_{i\beta} e_{i\gamma} f_i^{eq}$$

$$= \sum_i 3\rho c^2 w_i e_{i\alpha} e_{i\beta} e_{i\gamma} e_{i\theta} u_\theta$$

$$= 3c^2 \rho \sum_i w_i e_{i\alpha} e_{i\beta} e_{i\gamma} e_{i\theta} u_\theta$$

$$= 3c^2 \rho \left[ \frac{1}{9} \times 2\delta_{\alpha\beta\gamma\theta} u_\theta + \frac{1}{36} \times 4 \left( \delta_{\alpha\beta}\delta_{\gamma\theta} + \delta_{\alpha\gamma}\delta_{\beta\theta} + \delta_{\alpha\theta}\delta_{\gamma\beta} \right) u_\theta - \frac{1}{36} \times 8\delta_{\alpha\beta\gamma\theta} u_\theta \right]$$

$$= \frac{1}{3} c^2 \rho \left( \delta_{\alpha\beta}\delta_{\gamma\theta} + \delta_{\alpha\gamma}\delta_{\beta\theta} + \delta_{\alpha\theta}\delta_{\gamma\beta} \right) u_\theta$$

$$= \rho \theta \left( u_\alpha \delta_{\beta\gamma} + u_\beta \delta_{\gamma\alpha} + u_\gamma \delta_{\alpha\beta} \right)$$

Then we obtain

$$\sum_i c_i^2 c_{i\alpha} f_i^{eq} = \sum_i c_{i\alpha} c_{i\beta} c_{i\beta} f_i^{eq} = (D+2) \rho \theta u_\alpha \qquad (A5)$$

Based on Eq. (A1)

$$(\partial_t + c_{i\alpha} \partial_\alpha) f_i^{eq} = -\omega f_i^{(1)}$$

$$\partial_t \left( \sum_i \frac{1}{2} c_i^2 f_i^{eq} \right) + \partial_\alpha \left( \sum_i \frac{1}{2} c_i^2 c_{i\alpha} f_i^{eq} \right) = -\omega \sum_i \frac{1}{2} c_i^2 f_i^{(1)}$$

Based on Eq. (A3), (A4), (A5)

$$\partial_{t_1} \left( \frac{D}{2} \rho \theta + \frac{1}{2} \rho u^2 \right) + \partial_{1\alpha} \left( \frac{D+2}{2} \rho \theta u_\alpha \right) = 0 \qquad (A6)$$

$\varepsilon \times$ Eq. (A6), then Eq. (23) can be obtained.

$$\partial_t \left( \frac{D}{2} \rho \theta + \frac{1}{2} \rho u^2 \right) + \partial_\alpha \left( \frac{D+2}{2} \rho \theta u_\alpha \right) = 0$$